\DocumentMetadata{}
\documentclass[sigconf]{acmart}

\usepackage{xspace}
\usepackage{enumitem}
\usepackage{tcolorbox}
\usepackage{multirow}
\usepackage{multicol}
\usepackage{amsmath,amsfonts}
\usepackage{algorithm}
\usepackage{algorithmic}
\usepackage{hyperref}
\usepackage{hyperxmp}
\usepackage{graphicx}
\usepackage{subcaption}
\usepackage{wrapfig}
\usepackage{rotating}

\usepackage{xcolor}

\usepackage{listings}
\usepackage{tablefootnote}

\usepackage{soul}
\usepackage{xcolor}
\sethlcolor{gray!20}

\newcommand{\lstbg}[3][0pt]{{\fboxsep#1\colorbox{#2}{\strut #3}}}
\lstdefinelanguage{diff}{
  basicstyle=\ttfamily\footnotesize,
  morecomment=[f][\lstbg{red!20}]-,
  morecomment=[f][\lstbg{green!20}]+,
  morecomment=[f][\textit]{@@},
}

\newcommand{\find}[1]{
}

\AtBeginDocument{%
  }

\setcopyright{acmcopyright}
\copyrightyear{2025}
\acmYear{2025}
\acmDOI{XXXXXXX.XXXXXXX}

\acmConference[ICSE 2026]{48th IEEE/ACM International Conference on Software Engineering}{12 - 18 April 2026}{Rio De Janeiro, Brazil}

\newboolean{showcomments}
\setboolean{showcomments}{false}
\ifthenelse{\boolean{showcomments}}
{ }
\ifthenelse{\boolean{showcomments}}
{ }

\newcommand{\thanh}[1]{}

\newcommand*{\ours}{\textsc{FLAMES}\@\xspace}

\begin{document}
%%
%% The "title" command has an optional parameter,
%% allowing the author to define a "short title" to be used in page headers.

\title[Memory-Efficient LLMs for Program Repair with Semantic-Guided Patch Generation]{Memory-Efficient Large Language Models for Program Repair with Semantic-Guided Patch Generation}

% \title[Semantic-guided Search for Efficient Program Repair with LLMs]{Semantic-guided Search for Efficient Program Repair with Large Language Models}

\author{Thanh Le-Cong}
\email{congthanh.le@student.unimelb.edu.au}
\affiliation{%
	\institution{The University of Melbourne}
	\city{Melbourne}
	\country{Australia}
}

\author{Bach Le}
\email{bach.le@unimelb.edu.au}
\affiliation{%
	\institution{The University of Melbourne}
	\city{Melbourne}
	\country{Australia}
}

\author{Toby Murray}
\email{toby.murray@student.unimelb.edu.au}
\affiliation{%
	\institution{The University of Melbourne}
	\city{Melbourne}
	\country{Australia}
}

\begin{abstract}
Fixing software bugs is crucial yet demands significant resources from developers. Automated Program Repair (APR) is a promising solution to address this challenging task. The emergence of Large Language Models (LLMs) has opened a new era of LLM-based APR, substantially advancing the APR field further. LLM-based APR methods face significant challenges regarding memory inefficiency, hindering their scalability and effectiveness. This is largely due to the beam search utilized in the patch generation phase of LLM-based APR, which requires large beam sizes to search for more potentially good repair candidates. 

In this paper, we first show that increases in beam size, even for small-sized LLMs (1B-7B params), require extensive GPU usage, leading to up to 80\% of recurring crashes due to memory overloads in LLM-based APR. Seemingly simple solutions to reduce memory consumption are (1) to quantize LLM models, i.e., converting the weights of an LLM from high-precision values to lower-precision ones, and (2) to make beam search sequential, i.e., forwarding each beam through the model sequentially and then concatenating them back into a single output. However, we show that these approaches still do not work via both theoretical analysis and experiments. 

To address this, we introduce \ours, a novel LLM-based APR technique that employs semantic-guided patch generation to enhance repair effectiveness and memory efficiency. Unlike conventional methods that rely on beam search, \ours utilizes greedy decoding to enhance memory efficiency while steering the search towards more potentially good repair candidates via a semantic-guided best-first search algorithm. At each decoding step, \ours uses semantic feedback from test validation, such as the number of passing and failing test cases, to select the most promising token to explore further. Our empirical evaluation on Defects4J shows that \ours substantially reduces memory consumption by up to 83\% compared to LLM-based APR without compromising time efficiency. Moreover, \ours correctly fixes 133 bugs on Defects4J, fixing 10 bugs more than the best baseline. Additionally, these improvements also generalize to the HumanEval-Java and TransformedD4J datasets, where \ours generates 12\% and 36.5\% more correct patches, respectively, than the best baseline.

\end{abstract}


% \ccsdesc[500]{Computing methodologies~Natural language processing}


\maketitle

\section{Introduction}~\label{sec:Introduction}

Fixing software bugs is a complex and time-consuming task for developers~\cite{winter2022developers}. Automated Program Repair (APR)~\cite{goues2019automated, gao2022program} has emerged as a promising solution to alleviate this burden by automatically repairing bugs. Over the past two decades, APR has seen significant advancements~\cite{le2016history, le2017s3, liu2019tbar, ghanbari2019practical, chen2019sequencer,  le2021usability, zhu2021syntax, xia2022less, selfAPR2022}, with practical applications in the software industry~\cite{winter2022towards, kirbas2021introduction, marginean2019sapfix, bader2019getafix}. 

Recently, the surge of Large Language Models (LLMs), pre-trained on vast datasets, has further advanced APR, opening a new era of LLM-based APR~\cite{silva2023repairllama, xia2023aprllm, jiang2023impact}. LLM-based APR typically adapts models for APR either through fine-tuning on APR-specific datasets~\cite{silva2023repairllama, jiang2023impact, huang2023empirical, paul2023enhancing, zhang2022coditt5, zirak2024improving} or by using prompting with commercial models~\cite{bouzenia2024repairagent, yin2024thinkrepair}. In this study, we focus on fine-tuning approaches, as prompting often relies on closed-source and commercial models like ChatGPT, which lack transparency, hindering reproducibility in open science and raising data privacy concerns.

Fine-tuning-based approaches typically refine the weights of Code LLMs by fine-tuning these models on APR datasets. During inference, they generate token sequences to construct candidate patches, prioritizing sequences with high probability scores from the fine-tuned model. Traditionally, beam search, a widely used search algorithm from natural language processing~\cite{sutskever2014sequence}, is used to optimize this process by maintaining only the top-$n$ nodes at each decoding step, where $n$ is the beam size. A larger beam size provides a more precise approximation to exact decoding, i.e., thoroughly exploring all  possible sequences, thereby enhancing the quality of the outputs. This hyperparameter is even more important in the context of APR, which covers a broader range of possible identifiers and a larger search space than natural language~\cite{lutellier2020coconut,chen2019sequencer,jiang2021cure}. It has been shown that a substantial beam size is essential for generating adequate candidate patches for optimal results~\cite{jiang2021cure, lutellier2020coconut, zhong2024benchmarking}, and a larger beam size leads to higher performance of APR techniques that are based on small-sized deep learning models (less than 300M parameters) ~\cite{tufano2019empirical, ye2022neural, zhong2024benchmarking}. 

But, \textit{is it true that increasing the beam size is all we need to improve the performance of APR?} We revisit the impact of beam size on the effectiveness of LLM-based APR techniques on larger models (1B-7B parameters). Our findings reveal that increasing the beam size boosts the effectiveness of these techniques, but once the beam size reaches a threshold, the performance drops significantly due to memory overloads that cause recurring crashes even on state-of-the-art hardware (NVIDIA A100 GPU, equipped with 80 GB of VRAM). For instance, state-of-the-art LLM-based APR techniques generated between 21\% and 46\% more plausible patches when the beam size was increased from 10 to 25. Ideally, we expect that further increasing the beam size would lead to better performance of the models for APR. However, when the beam size is set higher at 50, 100, and 200, our experiments reveal that the performance of these models drops significantly.

Our experiments uncover that the above phenomenon is due to extensive GPU resource consumption, particularly on Video Random Access Memory (VRAM), caused by larger beam sizes. Often, upgrading VRAM to accommodate larger beam sizes requires GPU upgrades, which are expensive; thus, this creates a costly trade-off between memory efficiency and the performance of LLM-based APRs. For instance, achieving a 46\% performance improvement with InCoder-1B required 58\% more average and 111\% more peak VRAM usage on our dataset. Consequently, further performance gains through increasing beam size demand substantial VRAM, often leading to out-of-memory (OOM) crashes even on cutting-edge hardware and thereby unduly reducing the effectiveness of LLM-based APR techniques. Our experiments using the NVIDIA A100 GPU, equipped with 80 GB of VRAM, showed that InCoder-6B crashed on nearly 80\% of the evaluated bugs when using a beam size of 200. This resulted in a 60\% reduction in performance compared to a beam size of 10. In summary, we conclude that simply increasing the beam size is not a solution, as it comes at the cost of chasing after state-of-the-art hardware. 

A seemingly straightforward approach to this problem is to reduce the memory usage of LLM-based APR through engineering efforts. This can be achieved using two common strategies: (1) quantizing LLM models by converting their weights from high-precision to lower-precision values, and (2) making beam search sequential, where each beam is processed individually through the model before being combined into a single output. However, our findings show that while these methods can reduce memory usage, their memory demands still escalate significantly as beam size increases, leading to substantially high out-of-memory (OOM) rates. Consequently, these engineering efforts alone cannot solve the problem. This realization motivates us to develop a patch generation algorithm that can effectively enhance the performance of LLM-based APR techniques without relying on increasing the beam size and compromising memory efficiency.

In this paper, we introduce \ours, a memory-efficient LLM-based APR technique using semantic-guided patch generation. Unlike conventional methods that rely solely on a language model's knowledge and beam search, our approach aims to leverage semantic feedback from test validations to guide LLMs in patch generation. The core idea of \ours is to combine LLM-based and search-based APR. This process begins with generating initial patches using greedy decoding, i.e., beam search with a beam size of 1, followed by iterative, semantic-guided searches to refine these solutions. To achieve this, we employ a best-first search algorithm, specifically PG-TD~\cite{zhang2023planning}, along with semantic feedback from test validations to guide the patch generation of LLMs. This approach offers three key advantages: (1) reduced VRAM consumption through greedy decoding, improving memory efficiency; (2) scalability without increasing VRAM usage while generating more candidate patches; and (3) seamless information exchange between patch generation and validation, enabling efficient exploration of plausible patches.

We conducted an empirical evaluation of our proposed approach, \ours, using a dataset comprising 333 bugs from Defects4J~\cite{just2014defects4j}, 163 bugs from HumanEval-Java~\cite{jiang2021cure}, and 1,098 bugs from TransformedD4J~\cite{le2024evaluating}. We compared \ours against 15 state-of-the-art APR techniques and Qwen2.5-Coder-32B, the leading open-source LLM for code in well-known leaderboards~\cite{liu2023your, zhuo2024bigcodebench}. Our experimental results demonstrate that \ours can correctly fix 8\%, 12\%, and 36.5\% more bugs than the best baseline on Defects4J, HumanEval-Java, and TransformedD4J, respectively. 
We also found that \ours's semantic-guided patch generation substantially outperforms widely used patch generation strategies in LLM-based APR, including beam search and multiple sampling, with at least 12\% and 23\% improvements in the number of correctly fixed bugs.
Moreover, our analysis of memory efficiency showed that our method could reduce VRAM consumption by 42\% to 83\%, decreasing peak VRAM requirements from over 80 GB to as low as 12.7 GB across various configurations and models. Despite the improvement in memory efficiency, \ours does not compromise time efficiency, even generating plausible patches faster than conventional LLMs using beam search in many cases.

In summary, we have made the following contributions:

\begin{itemize}[leftmargin=*]
    \item We empirically study the impact of beam size on the effectiveness and memory efficiency of five different LLM-based APR techniques, highlighting the challenges of scaling the search space due to extensive memory consumption of the techniques.

    \item We introduce a novel approach, namely \ours, which fuses LLM-based and search-based APR, utilizing feedbacks from patch validation to efficiently discover plausible candidate patches. 
    
    \item We empirically evaluate the performance of \ours against 15 state-of-the-art baselines and the leading open-source LLM for code. \ours correctly fixes 133 bugs in Defects4J, 103 bugs in HumanEval-Java and 456 bugs in TransformedD4J, surpassing the best baseline by 8\%, 12\% and 36.5\%, respectively. \ours also outperforms widely-used patch generation strategies by at least 12\%. Moreover, \ours substantially reduces memory consumption by up to 83\% without compromising time efficiency.  
\end{itemize}

\section{Motivation Study}~\label{sec:motivation}

Beam size is a critical hyperparameter in beam search, the fundamental algorithm for patch generation in LLM-based APR. In this section, we evaluate the impact of beam size on the effectiveness and memory efficiency of LLM-based APR techniques using LLMs.  
Although prior studies~\cite{jiang2021cure, lutellier2020coconut, xia2022less, wei2023copiloting} showed that a substantial beam size is needed for optimal APR performance, our experiments reveal that increasing the beam size unduly hinders memory efficiency and subsequently degrades the performance of LLM-based APR, even on state-of-the-art hardware. 

\begin{table*}[htbp]
    \centering
    \begin{tabular}{l||c|c||c|c||c}
        \hline
        \textbf{Models} & \textbf{InCoder-1B} & \textbf{InCoder-6B} & \textbf{CodeGen-2B} & \textbf{CodeGen-6B} & \textbf{RepairLLama}\\
        \hline
        
        \textbf{Base Model} & \multicolumn{2}{c||}{InCoder~\cite{fried2022incoder}} & \multicolumn{2}{c||}{CodeGen-NL~\cite{nijkamp2022codegen}} & CodeLlama-7B~\cite{roziere2023code} \\
        \hline
        
        \textbf{Pre-training Dataset} & \multicolumn{2}{c||}{InCoder~\cite{fried2022incoder}} & \multicolumn{2}{c||}{ThePile~\cite{gao2020pile}} & CodeLLama~\cite{roziere2023code} \\

        \textbf{- Cut-off Date} & \multicolumn{2}{c||}{12.2021} & \multicolumn{2}{c||}{12.2020} & 08.2023 \\
        \hline
        
        \textbf{Fine-tuning Dataset} & \multicolumn{4}{c||}{Recoder~\cite{zhu2021syntax}} & MegaDiff~\cite{monperrus2021megadiff} \\

        \textbf{- Cut-off Date} & \multicolumn{4}{c||}{03.2018} & 09.2015 \\
        \hline
        
        \textbf{Fine-tuning Technique} & \multicolumn{4}{c||}{Full-parameter Fine-tuning} & QLORA~\cite{dettmers2024qlora} \\
        \hline
        
        \textbf{Model Size} & 1B & 6B & 2B & 6B & 7B \\
        \hline
    \end{tabular}
    \caption{Detailed Information of LLM-based Program Repair techniques used in this study.}
    \label{tab:llm_info}
    \vspace{-5mm}
\end{table*}

\textbf{RQ$_1$}: \textbf{How does beam size affect memory efficiency of the LLM-based APR?} We measure the use of VRAM and the frequency of out-of-memory crashes in the LLM-based APR techniques with different beam sizes to understand the impact of this parameter on memory efficiency.
We selected beam sizes of 10, 25, 50, 100, and 200, as these are common in prior works~\cite{ye2022neural, xia2023aprllm, xia2022less, zhu2021syntax}. We do not go beyond a beam size of 200 due to hardware constraints. 

\textbf{RQ$_2$}: \textbf{How does beam size affect the effectiveness of LLM-based APR?} We investigate the impact of beam size on the effectiveness of five LLM-based APR techniques by measuring the number of plausible patches. Similarly to RQ$_1$, we also selected beam sizes of 10, 25, 50, 100, and 200. 

\subsection{Experimental Design}~\label{sec:motivation_design} 

\subsubsection{Benchmark Dataset.} To address these research questions, we conducted experiments with Defects4J, a benchmark of 835 real-world bugs from 17 open-source Java projects~\cite{just2014defects4j}. Following prior studies~\cite{ye2022neural, le2024evaluating, xia2022less, liu2019tbar, chen2019sequencer}, we focused on 333 single-hunk bugs, where patches modify a single contiguous code chunk. This selection aligns with LLM-based APR techniques like InCoder and CodeGen, which are fine-tune for single-hunk bug fixes~\cite{jiang2023impact}.

\subsubsection{Studied LLM-based APR techniques}~\label{sec:llm} To select the target techniques for our empirical study, we conducted a literature review and identified suitable techniques based on the following criteria: (1) use beam search as their patch generation strategy, (2) employ LLM with at least one billion parameters, and (3) provide accessible fine-tuned models for Automated Program Repair. 
Applying these criteria, we identified five LLM-based APR techniques from recent works~\cite{jiang2023impact, silva2023repairllama}: CodeGen (2B, 6B) and InCoder (1B, 6B), fine-tuned using full-parameter fine-tuning by Jiang et al.\cite{jiang2023impact}; and RepairLlama (7B), fine-tuned with efficient-parameter fine-tuning, i.e., QLORA, by Silva et al.~\cite{silva2023repairllama}. Detailed information about these LLMs are provided in Table~\ref{tab:llm_info}.
Since RepairLlama has utilized quantization, we excluded its quantized versions from follow-up experiments.

\subsubsection{Memory Reduction techniques.}~\label{sec:memory_reduction} In this study, we also investigate the potential of memory reduction techniques, including quantization and sequential beam search, through engineering efforts and their impact on our findings. \textbf{Quantization} is a widely used technique to reduce the memory usage of LLMs by converting their weights from high- to lower-precision values. In this work, we applied 4-bit quantization, reducing the LLM's weights from 32 bits to 4 bits. 
For clarity, we denote the original LLM-based APR as \textbf{full-precision}, and its quantized counterpart as \textbf{quantized} in the following sections.
Meanwhile, \textbf{sequential beam search} is a variant of the implementation of the beam search proposed by Saibo Geng~\cite{sequentialbs}, which processes each beam sequentially through the model before concatenating them into a single output. 

\subsubsection{Implementation Details.} We implemented LLM-based APR techniques in Python using PyTorch and HuggingFace. To improve reliability and reduce bias, we integrated inference code, repair prompts, and trained models from the original repositories into a unified framework. Our implementations also used HuggingFace's standard beam search. Inference was performed with a batch size of one, processing one buggy program at a time. All experiments were run on a single NVIDIA A100 GPU (80GB VRAM).

\subsection{\textbf{RQ$_1$}: 
Impact of Beam Size on Memory Efficiency}~\label{sec:rq1}

\begin{figure*}
    \centering
    \begin{subfigure}[b]{0.48\textwidth}
        \centering
        \includegraphics[width=\textwidth]{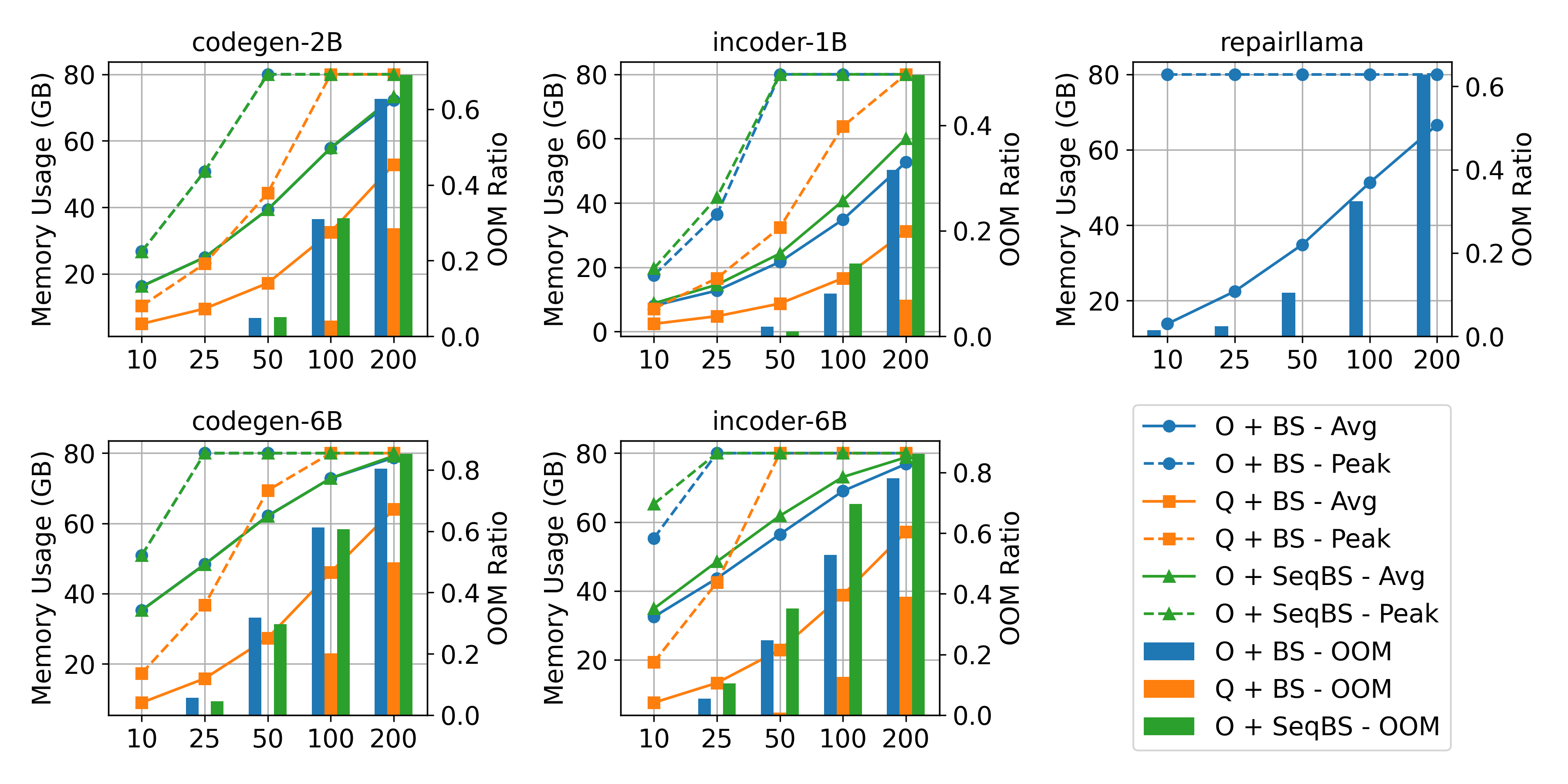}
        \caption{Memory Usage}
        \label{fig:llm_efficiency}
    \end{subfigure}
    \hfill
    \begin{subfigure}[b]{0.48\textwidth}
        \centering
        \includegraphics[width=\textwidth]{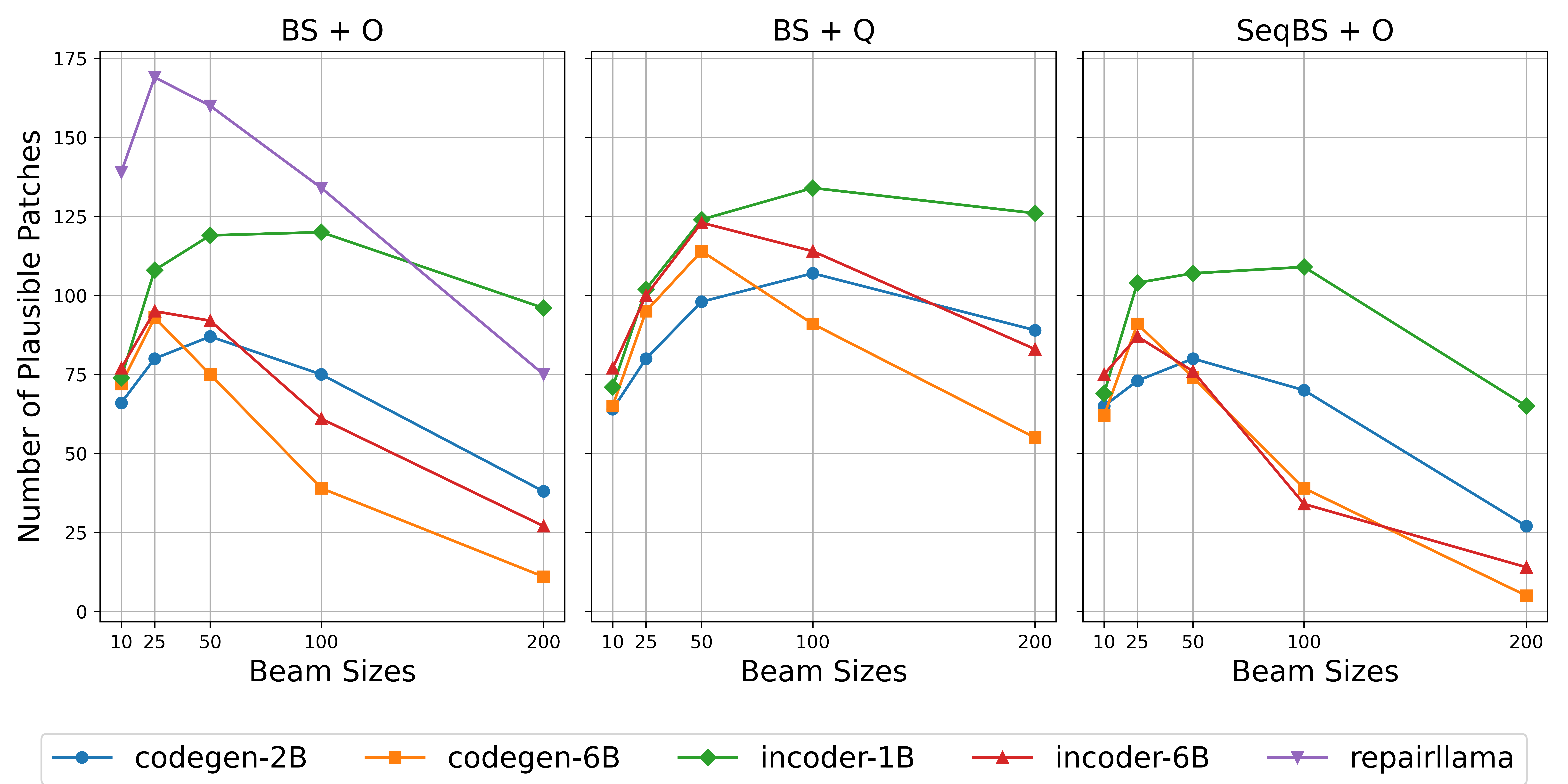}
        \caption{Effectiveness}
        \label{fig:llm_effectiveness}
    \end{subfigure}
    
    \caption{Memory usage and effectiveness of LLM-based APR techniques across beam sizes on an NVIDIA A100 (80GB) using full-precision (FP) and quantized (Q) models. BS and SeqBS denote standard and sequential beam search.}
    \label{fig:llm_performance_analysis}
\end{figure*}

Figure~\ref{fig:llm_efficiency} illustrates the experimental results of RQ$_{1}$. Overall, we can see that \textbf{VRAM usage increases significantly as the beam size increases from 0 to 200 in the full-precision versions} of five studied techniques. For instance, in InCoder-1B, the average memory usage substantially increases from 20GB at a beam size of 10 to around 60GB at a beam size of 200. Similarly, peak memory usage increases dramatically, reaching maximum memory limits at a beam size of only 50. Notably, this increase in VRAM usage leads to \textbf{a substantial increase in crashes due to Out-of-Memory (OOM) errors}. For example, the OOM ratio in InCoder-1B increases from nearly negligible at smaller beam sizes to approximately 50\% at a beam size of 200. Larger models, e.g. CodeGen-6B and InCoder-6B, show even higher OOM rates, nearly 80\%, respectively.

Next, we investigated 4-bit quantization and sequential beam search, commonly used to optimize model memory requirements, for reducing VRAM usage. As shown in Figure~\ref{fig:llm_efficiency}, quantization significantly reduces the average and peak VRAM consumption, thereby reducing OOM crashes. However, sequential beam search presents trade-offs and does not consistently decrease memory usage, as confirmed by its authors and our theoretical analysis (for details, please see our online Appendix~\cite{package}). More importantly, \textbf{even with these techniques, VRAM usage for LLM-based APR techniques continues to increase with increasing beam size}. For example, CodeGen-2B's average VRAM usage rises from less than 10GB at a beam size of 10 to 60GB at 200, with peak usage nearing the 80GB hardware limit. As a result, despite memory reduction techniques, LLM-based APR techniques still face frequent OOM crashes, occurring in 30–85\% of cases with a beam size of 200.

\begin{tcolorbox}
    \underline{\textbf{Answers to RQ$_1$:}} Both the average and peak memory usage of LLM-based APR techniques increase significantly as the beam size increases, inducing up to 80\% out-of-memory crashes on an A100 with 80GB of VRAM. Even with memory reduction techniques, memory usage still escalates, and crash rates remain substantial as the beam size increases.
\end{tcolorbox}

\subsection{RQ$_2$: Impact of Beam Size on Effectiveness}~\label{sec:rq2}

Figure~\ref{fig:llm_effectiveness} presents our experimental results for RQ$_{2}$. Overall, we can see that \textbf{the number of plausible patches generated by full-precision versions of LLM-based APR only shows an upward trend at smaller beam sizes and remarkably declines at larger sizes}. For instance, the number of plausible patches generated by RepairLlama increases from 139 at a beam size of 10 to 169 at 25, then drops significantly to 134 at 100 and further to 75 at 200, although the ratio of plausible patches remains at 0.60. This reduction is primarily due to the high frequency of OOM crashes, which range from 30-60\% at larger beam sizes, as depicted in Figure~\ref{fig:llm_efficiency}. These results highlight the significant impact of LLMs' extensive memory usage on their performance with large beam sizes.

Next, similar to RQ$_{1}$, we examined the impact of 4-bit quantization and sequential beam search on memory usage and the effectiveness of LLM-based APR techniques. \textbf{Overall, the effectiveness of LLM-based APR techniques remains consistent with sequential beam search, maintaining the same trends. Meanwhile, quantization maintains an upward trend in effectiveness up to a beam size of 100}. For example, the number of plausible patches generated by quantized CodeGen-2B increased from 64 at a beam size of 10 to 107 at a beam size of 100, surpassing the original full-precision model by 42.7\% at the same beam size. This improvement is due to a reduction in OOM crashes, from 30\% to less than 5\%. However, despite these gains, the effectiveness of quantized models still declines at larger beam sizes, such as 200, similar to the trend seen in the full-precision models. For example, the effectiveness of the quantized CodeGen-2B dropped from 114 plausible patches at a beam size of 50 to 91 at 100, and further to 55 at 200, reflecting declines of 20\% and 48\%, respectively.        

\begin{tcolorbox}
    \underline{\textbf{Answers to RQ$_2$:}} As the beam size increases, the effectiveness of LLM-based APR increases, peaking at mid-range beam sizes before declining at larger sizes due to higher OOM crash rates, particularly in larger models. Even with memory reduction techniques, the APR's effectiveness still quickly degrades at beam sizes larger than 100.
\end{tcolorbox}

\section{Approach}\label{sec:approach}

Our findings in Section~\ref{sec:motivation} highlight that scaling the beam size to a certain threshold may help improve the effectiveness of APR at the cost of extensive memory consumption, above which the effectiveness quickly degrades even on state-of-the-art hardware. To address these limitations, we propose a novel approach, \ours, which combines LLM-based and search-based APR for memory-efficient, semantic-guided patch generation.

\begin{figure}
    \centering
    \includegraphics[width=\columnwidth]{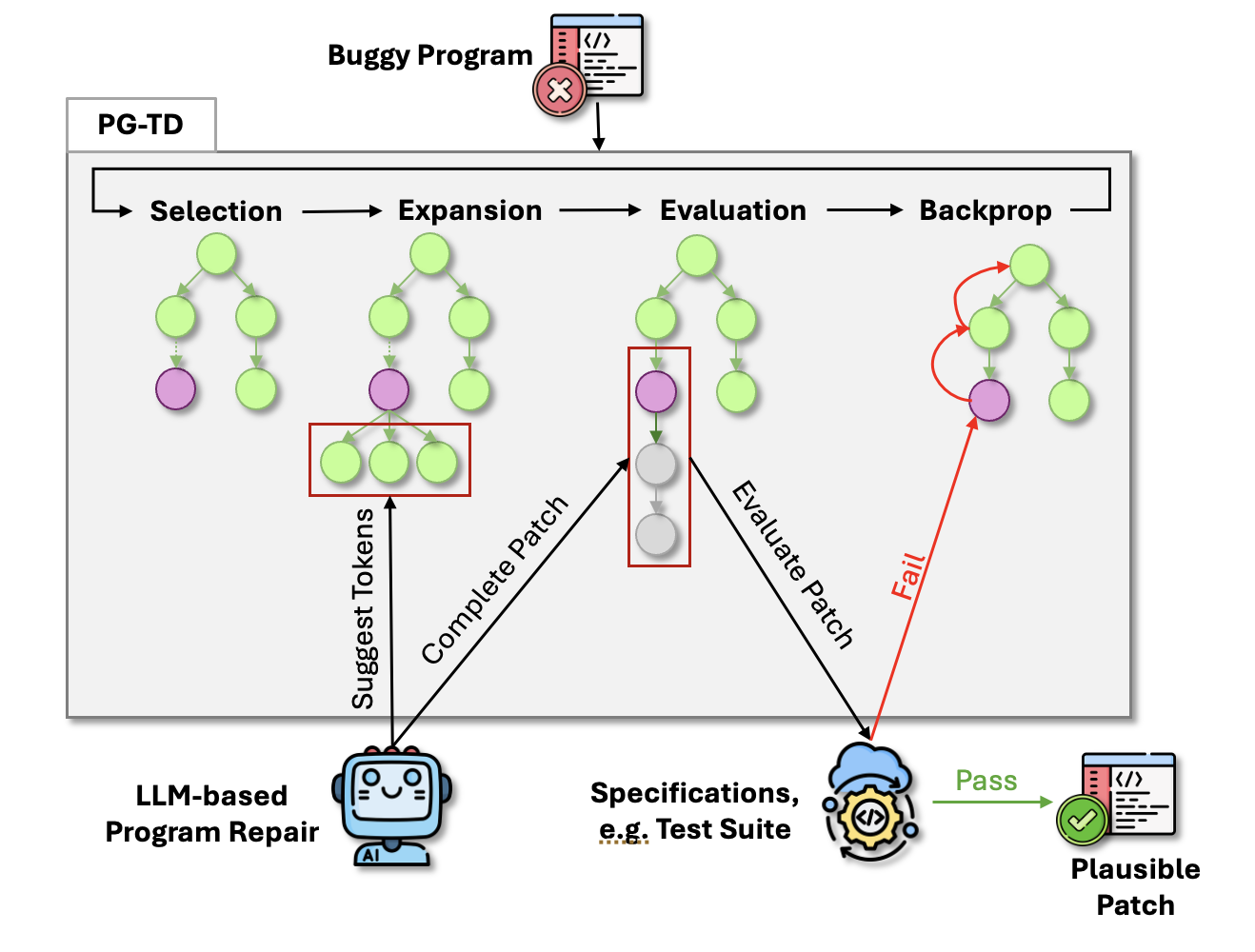}
    \caption{Overview of \ours}
    \label{fig:overview}
    \vspace{-3mm}

\end{figure}

Figure~\ref{fig:overview} illustrates the overall pipeline of \ours, which employs PG-TD as the core algorithm to guide patch generation in LLM-based APR. PG-TD is a best-first search algorithm, inspired by Monte Carlo Tree Search~\cite{kocsis2006bandit}, to decode Transformer models. PG-TD formulates the decoding process of Transformers models as a search problem on a tree structure whose nodes represent tokens. PG-TD then iteratively expands this tree with the information provided by the models and searches for an optimal candidate using a predefined reward function. 
In the following sections, we first present a high-level overview of patch generation of LLM-based APR with PG-TD and then present a detailed methodology of PG-TD's steps in APR context. 

\subsection{Patch Generation with PG-TD}

Algorithm~\ref{alg:overview} outlines the patch generation process in APR using the PG-TD algorithm. At a high level, PG-TD conducts an iterative search for plausible patches. This search proceeds by refining partial repair candidates over multiple iterations.
At each iteration, PG-TD selects the most promising partial patch from the current search space, as described in Section~\ref{sec:selection}. The chosen partial patch is then used in two key operations. First, since it is the most promising candidate, PG-TD will further expand the search space by incorporating most of the next potential tokens for the partial patch, predicted by an LLM-based APR model (Section~\ref{sec:extension}). Second, the partial patch is simulated into a complete patch candidate using the same LLM-based APR model and is subsequently evaluated against a predefined set of correctness specifications to compute a reward value (Section~\ref{sec:reward}). This reward is then backpropagated through the search tree to update the estimated potential of the nodes in the partial patch (Section~\ref{sec:backpropagation}). This iterative process continues until a plausible patch is discovered, that is, a candidate that satisfies the given specifications or the iteration limit is reached.

\subsection{Selection}~\label{sec:selection}

At this stage, our goal is to select the most promising partial patch to explore. To achieve this, we employ a policy algorithm that strikes a balance between exploiting potentially high-reward states and exploring less-visited ones. Specifically, an action (i.e., adding a token) is chosen based on three criteria: (1) the action's historical reward in relation to the current state is high, indicating strong potential; (2) language models (LLMs) predict that the action is a suitable next token, as reflected by its prediction probability; and (3) the resulting state after taking the action is under-explored.

To implement this approach, we evaluate three commonly used policy algorithms in Monte Carlo Tree Search (MCTS): UCB \cite{kocsis2006bandit}, and two variants of P-UCB \cite{silver2017mastering}, one with fixed weights (Fixed P-UCB) and the other with variable weights (Variable P-UCB). Our results indicate that Variable P-UCB is the most effective algorithm for the APR task, leading us to adopt it as our primary policy algorithm (See Section~\ref{sec:configuration} for further details.) 
 
\subsection{Expansion}~\label{sec:extension}

This stage aims to expand the search tree by adding possible next tokens. As the number of possible tokens in the vocabulary is huge, random sampling, which is used in MCTS~\cite{kocsis2006bandit} is likely to return in invalid and low-potential tokens. Therefore, \ours leverages fine-tuned LLM to suggest top-$k$ next tokens with the highest prediction score given the current state following PG-TD~\cite{zhang2023planning}. Note that, as a state can be re-visited multiple times during the tree search, \ours also leverages a caching mechanism to avoid redundant computation. In particular, whenever LLMs make a prediction, the input sequence, i.e., state and top-$k$ next tokens, is cached using a hash map. Then, when the algorithm needs LLM to predict top-$k$ next tokens, such information can be reused if it exists in the cache. In this work, we set $k=10$ and further discuss the impact of $k$ on our method in Sections~\ref{sec:k}.

\begin{algorithm}[t]
\caption{Semantic-guided Patch Generation with PG-TD. $SELECT$, $EXPAND$ and $BACKPROP$ are presented in Section~\ref{sec:selection}, ~\ref{sec:extension} and ~\ref{sec:backpropagation}, respectively. $SIMULATE$ and $EVAL$ are presented in Section~\ref{sec:reward}.}~\label{alg:overview}
\begin{algorithmic}[1]
\REQUIRE $bp$: buggy program; $\mathcal{L}$: LLM-based Program Repair model; $\mathcal{S}$: set of correctness specifications; $max\_iters$: number of iterative refinement steps
\ENSURE $cp$: plausible patch for $bp$ \OR $\mathit{None}$ if no patch found
\STATE $root \leftarrow \mathcal{L}.initial\_token$ \hfill $\triangleright$ Initialize search tree
\STATE $reward \leftarrow \{\}$ \hfill $\triangleright$ Initialize reward dictionary

\FOR{$i = 1$ \TO $max\_iters$}
    \STATE $pp \leftarrow SELECT(root, reward)$ \hfill $\triangleright$ Select best partial patch
    \STATE $root \leftarrow EXPAND(root, \mathcal{L}, bp)$ \hfill $\triangleright$ Expand search tree
    \STATE $cp \leftarrow SIMULATE(pp, \mathcal{L}, bp)$ \hfill $\triangleright$ Simulate complete patch $cp$
    \STATE $r \leftarrow EVAL(cp, \mathcal{S}, bp)$ \hfill $\triangleright$ Evaluate rewards of $pp$ using $cp$
    \IF{$r \neq 1$}
        \STATE $reward \leftarrow BACKPROP(reward, pp, r)$ \hfill $\triangleright$ Update rewards
    \ELSE
        \RETURN $cp$ \hfill $\triangleright$ Return plausible complete patch
    \ENDIF
\ENDFOR
\RETURN $\mathit{None}$
\end{algorithmic}
\end{algorithm}

\subsection{Evaluation}~\label{sec:reward} 

This stage aims to evaluate the potential of the current state, i.e., how likely the partial patch leads to a optimal (complete) patch. Similar to expansion, it is impossible to evaluate the current state using an random simulation as used in MCTS algorithm. Consequently, \ours leverages beam search to automatically complete the current state, i.e., a partial patch $pp$, to form a complete patch $cp$. However, it is noteworthy that leveraging the beam search as in original PG-TD still hinders memory efficiency. Therefore, instead of beam search, \ours utilizes a greedy search that is equivalent to a beam search with a beam of size 1, to find the complete patch for the current state while keeping low memory consumption. 

Then, given the completed patch, \ours estimates the potential of current state by calculating a reward function $R(cp, spec)$. Theoretically, $spec$ and reward function $R$ can takes any kinds of specification and a corresponding pre-defined reward function. In this work, following common practices in literature~\cite{gao2022program}, we leverage developer-written test cases as our specification and utilize the passing ratio, which is widely-used in search based APR~\cite{gao2022program}, as our reward function. More formally, given a program $cp$ and a set of test cases $spec$, we define our reward function as follows:
$R(cp, spec) = \frac{f_{pass}}{f_{pass} + f_{fail}}$ with $f_{pass}$ and $f_{fail}$ is the number of passing and failing test cases observed when running $cp$ on $spec$. 

\subsection{Backpropagation}~\label{sec:backpropagation}

This stage aims to update the value of nodes in the tree for guiding next steps of the search process. Intuitively, the stage allows \ours to provide feedback from test validation for guiding the patch generation process of LLMs. Particularly, the reward of a state (calculated based on automatically-completed program) is backpropagated to its parents recursively until the root is reached. 
Similar to other reinforcement learning algorithms~\cite{sutton2018reinforcement, wiering2012reinforcement, kaelbling1996reinforcement}, we aim to maintain the maximum observed reward as an expected return value for a given state $s$ and the corresponding action $a$.

To do so, for all state and action pair $(s, a)$ along the path from the current state to the root node, \ours updates their Q-value by $Q(s, a) = max(Q(s, a), R(cp, spec))$, where $R(cp, spec)$ is the reward value of the current state action pair $(s, a)$ with $cp$ is an automatically complete patch of $s$.

\subsection{Implementation Details.} ~\label{sec:implement}

We implemented \ours by extending the PG-TD~\cite{zhang2023planning} framework based on the DynaGym toolkit. We use RepairLLama, the best-performing model among LLM-based repair techniques, as our default LLM-based repair model. For repair, \ours generates up to 200 patches and sets a 30-minute timeout for each bug, as recommended by developers~\cite{noller2022trust}. Experiments are conducted on an NVIDIA A100 GPU with 80 GB VRAM, 250 GB RAM, and a 32-core Intel Xeon CPU at 2.90 GHz, running Red Hat Enterprise Linux 9.3 and Java 1.8.0\_241. This setup is comparable to those used in related works~\cite{silva2023repairllama, wei2023copiloting}.

\section{Empirical Evaluation}

\subsection{Research Questions}
Continuing from our motivation study in Section~\ref{sec:motivation}, our empirical evaluation aims to answer the following four research questions. 

\textbf{RQ$_{3}$}: \textbf{How effective is \ours?} We assess the effectiveness of \ours on the Defects4J~\cite{just2014defects4j}, HumanEval-Java~\cite{jiang2023impact} and TransformedD4J~\cite{le2024evaluating} datasets, comparing our proposed technique to 15 state-of-the-art APR techniques and Qwen2.5-Coder-32B, the best open-source LLM for code in well-known leaderboards~\cite{zhuo2024bigcodebench, liu2023your}.

\textbf{RQ$_{4}$}: \textbf{How efficient is \ours?} We evaluate the time and memory efficiency of \ours and compare it with standard and sequential beam search algorithms.

\textbf{RQ$_{5}$}: \textbf{How effective is \ours's semantic-guided patch generation?} We investigate the effectiveness of \ours's semantic-guided patch generation by implementing our approach with six LLMs for APR and comparing their effectiveness to the full-precision models using two widely used patch generation strategies: beam search~\cite{ye2022neural, silva2023repairllama} and multiple sampling~\cite{wei2023copiloting,xia2023revisiting}.

\textbf{RQ$_{6}$}: \textbf{How do different configurations affect the performance of \ours?} Finally, we examine the impact of various key factors on the effectiveness of \ours, including the reward function, the expansion size, and the policy algorithm.

\subsection{Experimental Setup}~\label{sec:setup}

\subsubsection{Benchmark Dataset.}~\label{sec:dataset} Similar to our initial motivation study (Section~\ref{sec:motivation_design}), we evaluated our method using 333 single-hunk bugs from the Defects4J dataset~\cite{just2014defects4j}. Furthermore, we used HumanEval-Java~\cite{jiang2023impact} and TransformedD4J~\cite{le2024evaluating}, which contain artificial bugs introduced by applying bug injection to normal programs and semantic-preserving transformations to Defects4J bugs, respectively, to mitigate the risk of data leakage.

\subsubsection{APR Baselines.} To evaluate the effectiveness of our approach (RQ3), we compared \ours with the leading baselines in various APR categories: Template-based, NMT-based, Cloze-based, and LLM-based APR techniques. For LLM-based APR, we evaluated five models using their best variants identified in RQ$_{2}$. For Cloze-based APR, we included FitRepair~\cite{xia2023revisiting}, AlphaRepair~\cite{xia2022less}, and Repilot~\cite{wei2023copiloting}. We also compared \ours with NMT-based methods, i.e., ITER~\cite{ye2024iter}, Recoder~\cite{zhu2021syntax}, KNOD~\cite{jiang2023knod}, and RewardRepair~\cite{ye2022neural}, as well as template-based approaches, i.e., GAMMA~\cite{zhang2023gamma}, TENURE~\cite{meng2023template}, and TBar~\cite{liu2019tbar}. Finally, we also included Qwen2.5-Coder-32B~\cite{hui2024qwen2}, the leading LLM for code on well-known leaderboards~\cite{liu2023your, zhuo2024bigcodebench}.
Following standard practices~\cite{jiang2023impact, jiang2021cure, chen2019sequencer, wei2023copiloting, xia2022less}, we performed evaluations under the assumption of perfect fault localization, which ensured that variations in fault localization techniques do not affect the results. For LLM-based techniques, we used the same configurations as in Section~\ref{sec:implement}. For other techniques, we collected bug fix results from their original papers and replication packages, following common practices in prior works~\cite{jiang2021cure, wei2023copiloting, xia2022less, ye2022neural}.

\subsubsection{Patch Generation Algorithm Baselines.} To evaluate the effectiveness of \ours's semantic-guided patch generation (RQ5), we also compared our approach with two well-known patch generation strategies: beam search and multiple sampling. For beam search, we leverage the standard implementations from the Huggingface library, a widely adopted toolset for training and deploying LLMs. For multiple sampling, we first set the LLM temperature to 1 to maximize its diversity. Then, we repeat the default patch generation of these models (using beam search) until we reach 200 patches. 

\subsubsection{Evaluation Metrics and Patch Correctness Assessment.} We assessed the effectiveness of APR techniques using a standard metric: the number of correct patches. To identify patch correctness, following prior works~\cite{ye2022neural, le2019reliability, xia2022less, lutellier2020coconut, le2024evaluating}, we conducted a manual evaluation to determine if they were syntactically or semantically equivalent to developer-written patches. Note that while manual patch correctness assessment is precise, it is resource-intensive and requires considerable human effort.

\subsection{RQ$_3$: Comparison with Existing Techniques}

\begin{table}
    \centering
    \caption{Comparison with state-of-the-art APR techniques on Defects4J benchmark. \#Correct and \#Plausible  denotes the number of correct and plausible patches generated by each tool. RepairLLama (+BS) and (+MS) denotes the results of RepairLLama with beam search and multiple sampling, respectively.}
    \label{tab:d4j_results}
    \begin{tabular}{lccc}
    \toprule
    \textbf{Techniques} & \textbf{\#Correct} & \textbf{\#Plausible} & \textbf{Precision} \\ \midrule
    \textbf{TBar} & 46 & - & - \\ 
    \textbf{GAMMA} & 87 & - & - \\
    \textbf{TENURE} & 84 & - & - \\ 
    \textbf{RewardRepair} & 79 & - & - \\
    \textbf{Recoder} & 43 & - & - \\
    \textbf{KNOD} & 110 & - & - \\
    \textbf{ITER} & 55 & - & - \\ 
    \textbf{AlphaRepair} & 95 & - & - \\
    \textbf{FitRepair} & 123 & - & - \\
    \textbf{Repilot} & 116 & - & - \\ 
    \textbf{InCoder-1B} & 42 & 75 & 0.56 \\
    \textbf{InCoder-6B} & 50 & 96 & 0.52 \\
    \textbf{CodeGen-2B} & 41 & 88 & 0.47 \\
    \textbf{CodeGen-6B} & 49 & 94 & 0.52 \\
    \textbf{Qwen2.5-Coder-32B} & 68 & 112 & 0.61 \\ 
    \textbf{RepairLLama (+BS)} & 112 & 170 & 0.66 \\ 
    \textbf{RepairLLama (+MS)} & 90 & 151 & 0.59 \\ 
    \textbf{FLAMES} & 133 & 211 & 0.63 \\ 
    \bottomrule
    \end{tabular}
    \vspace{-4mm}
\end{table}

\subsubsection{Repair Success.} Table~\ref{tab:d4j_results} presents the number of correct fixes generated by our approach, \ours, and baseline methods in Defects4J. \ours successfully repairs 133 of the 333 single-hunk bugs, outperforming all existing APR techniques. 

\textbf{Comparison with RepairLLama's Variants. }  Specifically, our approach generates correct patches for 21 and 43 more bugs than two other variants with beam search and multiple sampling of its core LLM, RepairLLama. These results highlight the advantages of our semantic-guided patch generation approach using PG-TD. This method enables \ours to scale to larger beam sizes without compromising memory efficiency, as observed in RepairLLama with beam search. Meanwhile, it also guides the patch generation process with semantic feedback, resulting in a more effective approach than multiple sampling without guidance.
Although this approach slightly reduces the precision from 0.66 to 0.63, this trade-off is justified by a 19\% increase in repair success. Moreover, \ours maintains a higher precision than other LLM-based APR techniques and RepairLLama with multiple sampling.

\textbf{Comparison with State-of-the-Art Techniques.} Additionally, \ours significantly outperforms other state-of-the-art APR baselines. It correctly fixes 10 more bugs than FitRepair, the best baseline, achieving an 8\% improvement. Additionally, \ours outperforms the top NMT-based technique, KNOD, and the top template-based approach, GAMMA, by 21\% and 53\%, respectively, in the number of correct patches. Notably, \ours achieves this performance with a budget of only 200 validated patches and a 30-minute pause per bug, while FitRepair and KNOD require 1,000–5,000 validated patches and a pause of 5-hours per bug.
Furthermore, \ours outperforms Qwen2.5-Coder-32B, the leading LLM for code, by 51\%. This advantage comes from two key factors: first, this model is not specifically trained for APR, unlike our core model, RepairLlama; second, \ours incorporates semantic-guided patch generation that enhances its repair effectiveness.

\textbf{Unique Fixes.} We found that \ours achieves the highest number of unique fixes among the best baselines for NMT-based (KNOD~\cite{jiang2023knod}), Cloze-based (FitRepair~\cite{xia2023revisiting}) and LLM-based (RepairLLama~\cite{silva2023repairllama}) APR approaches.
Particularly, RepairLlama, KNOD, and FitRepair contribute 6, 5, and 9 unique fixes, respectively, while FLAMES can fix 14 unique bugs. Moreover, there are 85 correct fixes that are not addressed by these four leading techniques. By combining all the techniques, we can successfully fix 299 bugs. This shows that \ours can be complementarily used with existing APR methods such as KNOD and FitRepair to significantly increase the number of correct fixes generated.

\begin{table}[]
    \centering
    \caption{Comparison with RepairLLama on multi-location bugs from Defects4J.}
    \begin{tabular}{lc}
         \hline
         \textbf{Techniques} & \textbf{Correct/Plausible Patches} \\ \hline
         \textbf{RepairLLama} & 16/35  \\
         \textbf{FLAMES} & 24/59 \\
         \hline
    \end{tabular}
    \label{tab:multi_location}
\end{table}

\subsubsection{Repair Success on Multi-Location Bugs.}~\label{sec:multi_locations} Beyond the comparison on single-location bugs, we also conduct experiments on multi-location bugs following the settings of RepairLLama~\cite{silva2023repairllama} with 172 bugs from Defects4J. In this setting, each bug requires edits on multiple non-consecutive locations within a single function. We compare the effectiveness of FLAMES with our base model and also the best LLM baseline, RepairLlama to further understand the impact of FLAMES on repairing multi-location bugs. The detailed results are presented in Table~\ref{tab:multi_location}.

Our experimental results show that while RepairLLama (the best LLM baseline) generates 35 plausible patches for these bugs, FLAMES (with RepairLLama as the base LLM) can generate 59 plausible patches, demonstrating 68\% improvement over the best baseline. Further inspection reveals that FLAMES generates 24 correct patches out of 59 plausible ones, compared to only 16 correct patches generated by RepairLLama. This result confirms the advantages of our patch generation for such complex scenarios.

\begin{table}
    \centering
    \caption{Comparison with state-of-the-art APR techniques on HumanEval-Java and TransformedD4J benchmark. The results are displayed as x/y, with x, y denotes the number of bugs with correct and plausible patches, respectively.}
    \label{tab:generalizability}
    \begin{tabular}{lcc}
    \toprule
    \textbf{Techniques} & \textbf{HumanEval-Java} & \textbf{TransformedD4J} \\ \midrule
    \textbf{CodeGen-6B} & 57/65 & \_/393 \\ 
    \textbf{InCoder-6B} & 71/81 & \_/427 \\ 
    \textbf{Qwen2.5-Coder-32B} & 68/111 & \_/403 \\ 
    \textbf{RepairLLama} & 92/118 & 334/581 \\ 
    \textbf{FLAMES} & 103/133 & 456/776 \\ 
    \bottomrule
    \end{tabular}
    \vspace{-4mm}
\end{table}

\subsubsection{Generalizability.} To address concerns related to data leakage~\cite{jiang2023impact} and benchmark overfitting~\cite{durieux2019empirical}, we extended our evaluation of \ours and the four most effective LLM-based APR techniques from the previous experiment. Specifically, we evaluated their performance on the HumanEval-Java~\cite{jiang2021cure} and TransformedD4J~\cite{le2024evaluating} benchmarks. Given the large number of plausible patches generated by these techniques on TransformedD4J, we limited our correctness assessment to patches produced by the two most effective methods, namely \ours and RepairLLama.

Overall, \ours significantly outperforms the baselines in terms of the number of correct patches. For example, in HumanEval-Java, \ours successfully fixes 11 more bugs than the best-performing baseline, RepairLlama, which generates 92 correct fixes, representing an improvement of 12\%. Similarly, in TransformedD4J, \ours achieves a 36.5\% improvement over the best baseline. These results demonstrate the generalizability of \ours across various evaluation benchmarks, highlighting its effectiveness in generating correct fixes in different data sets.

\begin{tcolorbox}
    \underline{\textbf{Answers to RQ$_3$ (Effectiveness):}} \ours successfully fixes 133, 103, and 456 bugs on Defects4J, HumanEval-Java, and TransformedD4J, significantly outperforming the best baselines by 8\%, 12\% and 36. 5\%, respectively.
\end{tcolorbox}

\subsection{RQ$_4$: Efficiency}~\label{sec:rq4}

\subsubsection{Memory efficiency.}
Table~\ref{tab:model_memory_usage} compares the memory usage of \ours with three baselines: (1) the full-precision LLM with beam search, (2) the quantized LLM with beam search, and (3) the full-precision LLM with sequential beam search, across five different LLMs. We evaluated VRAM usage using three key metrics: average VRAM usage, peak VRAM usage observed in our dataset, and out-of-memory ratios. Since our hardware is capped at 80GB VRAM, we assign a memory usage of 80GB to data points that experience OOM errors. Consequently, in cases of OOM, the reported memory usage represents a lower bound of the actual values.

\begin{table}
\centering
    \caption{Comparison of memory usage among the full-precision LLM (Full-Precision), quantized LLM (Quantized), full-precision LLM with sequential beam search (SeqBS) and full-precision LLM with \ours.}
    \label{tab:model_memory_usage}
 \begin{tabular}{lcccc}
    \toprule
        \textbf{Models} & \textbf{Method} & \textbf{Average} & \textbf{Peak} & \textbf{OOM}\\
            & & \textbf{(GB)} & \textbf{(GB)} & \textbf{(\%)} \\
        \midrule
          & \text{Full-Precision} & 72.2 & 80 & 62.8 \\
         \textbf{CodeGen-2B} & \text{Quantized} & 52.9 & 80 & 28.5\\
         & \text{SeqBS} & 68.2 & 80 & 69.1 \\
         & \text{PG-TD} & \textbf{12.1} & \textbf{13.9} & \textbf{0.0} \\ \hline
        & \text{Full-Precision} & 78.7 & 80 & 80.5 \\
         \textbf{CodeGen-6B} & \text{Quantized} & 64.0 & 80 & 49.8 \\
         & \text{SeqBS} & 75.5 & 80 & 85.3 \\
         & \text{PG-TD} & \textbf{27.1} & \textbf{31.7} & \textbf{0.0} \\ \hline
        & \text{Full-Precision} & 52.7 & 80 & 31.5 \\
         \textbf{InCoder-1B} & \text{Quantized} & 31.2 & 80 & 6.9 \\
         & \text{SeqBS} & 56.1 & 80 & 49.5 \\
         & \text{PG-TD} & \textbf{7.4} & \textbf{12.7} & \textbf{0.0} \\ \hline
          & \text{Full-Precision} & 76.8 & 80 & 78.1 \\
         \textbf{InCoder-6B} & \text{Quantized} & 57.2 & 80 & 39.0 \\
         & \text{SeqBS} & 75.6 & 80 & 86.6 \\
         & \text{PG-TD} & \textbf{32.4} & \textbf{37.4} & \textbf{0.0} \\ \hline
          & \text{Full-Precision} & 66.6 & 80 & 62.8 \\
         \textbf{RepairLlama} & \text{PG-TD} & \textbf{8.1} & \textbf{22.8} & \textbf{0.0} \\ \bottomrule
    \end{tabular}
    \vspace{-4mm}
\end{table}

Overall, the VRAM usage of \ours demonstrates a substantial reduction across all evaluated models, outperforming both the original full-precision configurations and the two widely used memory reduction techniques: Quantization and Sequential Beam Search (SeqBS). For example, in the case of the InCoder-1B model, the average VRAM usage drops from 52.7 GB in the original configuration to just 7.4 GB with \ours, representing a reduction of 86\%. In comparison, the quantization approach reduces memory usage to 31.2 GB, while SeqBS increases it to 56.1 GB, making \ours the most memory-efficient method.

Regarding peak VRAM usage, while baseline methods frequently reach the 80 GB VRAM limit, \ours maintains peak values well below this threshold. Notably, \ours achieves a maximum VRAM usage of just 12.7 GB for the smallest model (InCoder-1B) and 37.4 GB for the largest model (InCoder-6B). In contrast, both the original full-precision and SeqBS configurations consistently hit the 80 GB VRAM limits, increasing the risk of out-of-memory (OOM) errors.

Furthermore, in terms of OOM ratios, \ours completely eliminates OOM errors across all models, achieving a 0\% OOM ratio, while the original full-precision and SeqBS methods frequently experience severe memory failures. For example, the full-precision CodeGen-6B model encounters an 80.5\% OOM rate, making it \textbf{unusable} in memory-constrained environments. Even quantization, while reducing memory usage, does not guarantee the elimination of OOM errors, as observed with CodeGen-6B, which still reports an OOM ratio of 49.8\%.

\begin{tcolorbox}
    \underline{\textbf{Answers to RQ$_{4.1}$ (Memory Efficiency):}} \ours can significantly reduce VRAM usage in LLMs, successfully minimizing average VRAM usage by up to 83\% and reducing peak VRAM requirements from 80 GB to as low as 12.7 GB across various configurations and models. Additionally, \ours eliminates OOM errors, while the baselines witness OOMs in up to 86.6\% of the data points. 
\end{tcolorbox}

\begin{table}[]
    \centering
    \caption{Comparison of average running times of \ours and beam search (BS) for finding plausible patches with statistical measures on different models and number of generated patches (\# Patches)}
    \label{tab:running_time}
    \resizebox{\columnwidth}{!}{
    \begin{tabular}{lccccc}
    \toprule
    \textbf{Model} & \textbf{\# Patches} & \multicolumn{2}{c}{\textbf{Avg. Time (s)}} & $|\mathbf{\delta}|$ & \textbf{p-value} \\ \cline{3-4}
     & & \textbf{Flames} & \textbf{BS} & & \\ \midrule
    \multirow{5}{*}{\textbf{CodeGen-2B}} & 10 & \textbf{37.99} & 49.43 & 0.33 & 0.002 \\ 
     & 25 & \textbf{55.89} & 63.12 & 0.22 & 0.013 \\ 
     & 50 & \textbf{68.66} & 88.07 & 0.24 & 0.004 \\ 
     & 100 & \textbf{92.63} & 124.72 & 0.27 & 0.005 \\ 
     & 200 & \textbf{93.88} & 133.40 & 0.37 & 0.006 \\ \hline
    \multirow{5}{*}{\textbf{InCoder-1B}} & 10 & \textbf{48.18} & 68.71 & 0.35 & 0.001 \\ 
     & 25 & 91.39 & \textbf{86.82} & 0.07 & 0.195 \\ 
     & 50 & \textbf{115.79} & 138.61 & 0.11 & 0.081 \\ 
     & 100 & \textbf{129.62} & 177.19 & 0.12 & 0.059 \\ 
     & 200 & \textbf{161.58} & 199.17 & 0.08 & 0.178 \\ \hline
    \multirow{5}{*}{\textbf{RepairLLama}} & 10 & \textbf{44.40} & 57.55 & 0.25 & 0.001 \\ 
     & 25 & 90.95 & \textbf{79.21} & 0.13 & 0.027 \\ 
     & 50 & 103.79 & \textbf{96.81} & 0.15 & 0.014 \\ 
     & 100 & 126.74 & \textbf{111.42} & 0.15 & 0.022 \\ 
     & 200 & \textbf{98.42} & 101.35 & 0.22 & 0.018 \\ \bottomrule
    \end{tabular}
    }
    \vspace{-4mm}
\end{table}

\subsubsection{Time efficiency.} In this experiment, we evaluated the time efficiency of LLM-based APR using our proposed method, \ours, compared to traditional beam search techniques. Our aim is to validate the hypothesis that "\textit{\ours can identify plausible patches more quickly than the traditional beam search algorithm}". To validate this hypothesis, we measure the time each method takes to successfully identify plausible patches, employing the Mann-Whitney-Wilcoxon (MWW) statistical test and Cliff's Delta for effect size assessment. This experiment focuses on the CodeGen-2B, InCoder-1B, and RepairLlama models, chosen for their lower incidence of out-of-memory (OOM) crashes, ensuring a robust sample size for statistical testing. The detailed results are presented in Table~\ref{tab:running_time}.

For the CodeGen-2B model, our hypothesis is supported in all examined numbers of generated patches, with p values between 0.002 and 0.013 and effect sizes between 0.22 and 0.37, indicating that \ours is more time efficient than beam search. Notably, \ours reduce time by approximately 11.44 seconds and by 39.52 seconds with 10 and 200 generated patches.

The RepairLlama model also supports the hypothesis with p-values between 0.001 and 0.027 across all examined numbers of generated patches, confirming the time efficiency of \ours over the beam search, although the effect sizes range from negligible to small. Consequently, \ours shows lower average running times with 10 and 200 generated patches. For other configurations, while \ours generally performs faster than the beam search, the occasional longer run times of \ours skew the average, leading to variability in the results.

In contrast, the InCoder-1B model exhibits greater variability between the two methods. Although \ours tends to generate patches faster at higher numbers of generated patches (50, 100, and 200), it performs worse than BS with 25 generated patches. The most favorable performance for \ours is observed with 10 generated patches, where it significantly outperforms the beam search (p-value 0.001). However, for 25 and 200 generated patches, the differences in running times lack statistical significance, indicating that the benefits of \ours do not apply uniformly in all configurations within this model.

\begin{tcolorbox}
    \underline{\textbf{Answers to RQ$_{4.2}$ (Time Efficiency):}} \ours significantly improve time efficiency in LLMs, surpassing traditional beam search methods.
\end{tcolorbox}

\subsection{RQ$_5$: Effectiveness of Semantic Guided Patch Generation}~\label{sec:rq5}

In this experiment, we evaluated the effectiveness of \ours's semantic-guided patch generation in comparison to beam search and multiple sampling. These methods were evaluated in five LLM-based automated program repair (APR) models, as introduced in Section~\ref{sec:llm}, together with Qwen2.5-Coder-32B, a state-of-the-art LLM for code. The detailed results are presented in Table~\ref{tab:generalizability}.

\ours consistently outperforms beam search in all models, with improvement rates ranging from 12\% to 48\% in terms of the number of correct patches. For example, \ours generates 21 more correct patches than RepairLlama, our default base model. This is due to \ours’s effective use of GPU resources which allows for a more thorough exploration of the large patch space. In contrast, beam search requires substantial VRAM, resulting in a high out-of-memory crash rate, as discussed in RQ$_{4}$ and RQ$_{1}$.

\begin{table}
\centering
\caption{The number of plausible patches generated by \ours's patch generation algorithm compared to beam search (BS) and multiple sampling (MS) on different LLMs}~\label{tab:generalizability}
% \resizebox{0.45\textwidth}{!}{
\begin{tabular}{@{}lcccccc@{}}
\toprule
\textbf{LLM} & \textbf{BS} & \textbf{MS} & \textbf{\ours} \\ \midrule
\textbf{CodeGen-2B} & 41 & 38 & \textbf{57} \\
\textbf{InCoder-1B} & 57 & 52 & \textbf{64} \\
\textbf{RepairLLama} & 112 & 90 & \textbf{133} \\
\textbf{CodeGen-6B} & 49 & 39 & \textbf{70} \\
\textbf{InCoder-6B} & 50 & 44 & \textbf{74} \\ 
\textbf{Qwen2.5-Coder-32B} & \_ & 68 & \textbf{91} \\ 

\bottomrule
\end{tabular}
\vspace{-4mm}
\end{table}

A common approach to mitigate memory constraints is multiple sampling, which involves performing beam search with a low beam size and subsequently sampling the model multiple times to generate additional candidate patches. This method enables the generation of more candidate patches while reducing memory consumption. However, our findings indicate that this approach results in lower effectiveness compared to beam search. For instance, in the case of RepairLLama, multiple sampling leads to an approximately 20\% decline in performance. Consequently, \ours significantly outperforms multiple sampling across models, achieving improvements ranging from 23\% to 68\%. We hypothesize that this performance gap arises because \ours leverages feedback from test validation to guide patch generation, whereas multiple sampling relies solely on random sampling from the model’s output.

In conclusion, consistent performance improvements across various models strongly support the effectiveness of \ours's semantic-guided patch generation compared to widely used patch generation approaches such as beam search and multiple sampling.

\begin{tcolorbox}
\underline{\textbf{Answers to RQ$_{5}$ (Effectiveness of Patch Generation)}:} FLAMES's semantic-guided patch generation demonstrates superior performance across multiple LLMs in APR tasks, consistently outperforming widely used patch generation methods: beam search and multiple sampling, by at least 12\% and 23\%, respectively. 
\end{tcolorbox}

\subsection{RQ$_6$: Ablation study}~\label{sec:configuration}

To systematically assess how each component influences the performance of \ours, we investigated three key factors: the semantic-guided reward function (for the evaluation phase), the size of the expansion $k$ (for the expansion phase), and the choice of the policy algorithm (for the selection phase). 

% \begin{table}[h]
%     \centering
%     \caption{Impact of semantic reward function on \ours}
%     \label{tab:reward}
%     \resizebox{0.4\textwidth}{!}{
%     \begin{tabular}{lc}
%         \toprule
%         \textbf{Techniques} & \textbf{\# Plausible Patches} \\ \midrule 
%         \textbf{\ours} & 210  \\ 
%         \textbf{\ours}$_{\text{w/o semantic}}$ & 150 \\ \bottomrule
%     \end{tabular}
%     }
% \end{table}

\subsubsection{Reward Function.} In this experiment, we evaluate the effectiveness of our reward function, as detailed in Section~\ref{sec:reward}. We hypothesize that this function will guide the generation of more plausible patches by leveraging the ratio of passing test cases. To test this hypothesis, we compare our approach with a variant that depends solely on the models' probability outputs and disregards feedback from our semantic reward function. Our experimental results show that incorporating a semantic reward function can lead to a 40\% increase, from 150 to 210, in plausible patches. This substantial improvement highlights the value of semantic guidance in refining the search process to concentrate on patches that are more likely to succeed.

\subsubsection{Expansion Size ($k$).}~\label{sec:k} Next, we explore how the expansion size $k$ affects \ours by experimenting with various values of $k$, including 3, 5, 7, 10, 15, 20, 25 and 30. Our findings show that as $k$ increased from 3 to 10, the number of plausible patches increased from 190 to 210. Unfortunately, a further increase of $k$ to up to 30 showed minimal gains, with plausible patches ranging from 202 to 204. This is due to the broader, yet shallower search at higher k, which favors exploration over the exploitation needed for certain bugs.
This trend suggests that a larger $k$ may allow for a more thorough exploration of the solution space, enhancing the likelihood of identifying high-quality patches. However, this comes at a trade-off between the depth of exploration and exploitation. 

\subsubsection{Policy Algorithms.}
Finally, we investigated the impact of different policy algorithms on the generation of plausible patches. We examined the performance enhancement observed with advanced policy algorithms, including the original UCB \cite{kocsis2006bandit}, and two variants of P-UCB\cite{silver2017mastering}—one with fixed weights (Fixed P-UCB) and one with variable weights (Variable P-UCB). Our experimental results show that Variable P-UCB is the most effective algorithm, generating 210 plausible patches. This policy algorithm outperforms UCB and Fixed P-UCB by 17\% (179 plausible patches) and 11\% (188 plausible patches), respectively. The performance of Variable P-UCB likely stems from its ability to dynamically adjust the balance between exploration and exploitation during the search process, thereby optimizing the discovery of plausible patches.

\section{Discussion}
\subsection{Limitations}~\label{sec:limitations}

First, while our approach, \ours, correctly fixes 21 more bugs and 42 unique bugs compared to the original RepairLlama, there are 21 bugs that the original model could repair but \ours could not.
In an in-depth analysis on the 21 bugs that FLAMES cannot repair, we observed that the frequent pattern that these bugs most often require entirely new code segments. However, we also notice that these patterns appear in bugs that can be successfully repaired by FLAMES but cannot be repaired by RepairLlama. Therefore, it does not show a strong separation between the patterns of bugs fixed and not fixed by FLAMES. Therefore, we hypothesize that the failures stem from FLAMES’s trade-off strategy rather than specific types of bug. Specifically, FLAMES’s failure to repair these bugs can possibly be attributed to FLAMES’s smaller expansion size of 10. This limitation compromises \ours's exploration, resulting in a less comprehensive search space than that of RepairLlama's beam search, which considers up to 25 possible tokens of expansions. Consequently, \ours misses some bugs that could potentially be fixed with a broader search scope. Although a straightforward solution would be to increase the expansion size $k$, doing so introduces a trade-off between exploitation and exploration, as shown in Section~\ref{sec:k}. Therefore, further analysis is essential to optimize the balance between expansion size and efficiency to enhance \ours's performance.

Second, \ours is constrained by the design of its reward function, which relies on the number of passing and failing test cases. While our experiements demonstrated the effectiveness of our reward function on guiding LLM-based APR to generate more plausible patches and thereby more correct patches. It is important to note that a substantial increase in plausible patches does not always result in a proportionally large increase in correct patches.
Moreover, it also raise the problem of test overfitting as observed in search-based APR~\cite{qi2015analysis, le2018overfitting}. 
We believe that refining the reward function with additional criteria such as naturalness~\cite{xia2023automated, kang2022language}, syntax/semantic similarity~\cite{xin2017leveraging, wen2018context} or history versions~\cite{nguyen2024encoding, le2021refixar, tan2015relifix} could enhance both the effectiveness and precision of \ours.

\subsection{Threats to validity}

\subsubsection{Internal Validity.}  A main threat to internal validity stems from the issue of data leakage, where evaluation data might overlap with training and fine-tuning data. To mitigate this risk, we followed prior studies~\cite{silva2023repairllama, jiang2023impact} and utilized the HumanEval-Java dataset for evaluation. This dataset consists of 163 bugs introduced by injecting faults into the HumanEval benchmark~\cite{chen2021evaluating}. Additionally, we employed TransformedD4J, which comprises 1,090 bugs generated through semantic-preserving transformations of Defects4J, further reducing the likelihood of data leakage. Note that, HumanEval-Java and TransformedD4J datasets were publicly released in July 2023 and January 2024, respectively. In contrast, CodeLlama, the base LLM underlying RepairLlama, was released in August 2024. This temporal gap significantly reduces the likelihood of data leakage during the pre-training phase. Furthermore, the fine-tuning data used in the development of RepairLlama was extracted from the Boa dataset~\cite{dyer2013boa} collected in September 2015, further minimizing the possibility of overlap with our evaluation datasets. Therefore, we believe that the risk of data leakage in our study is negligible.
Another concern is the manual evaluation of patch correctness. To address this, authors and an external annotator rigorously analyze each patch to confirm semantic equivalence to developer-written patches. Moreover, to ensure transparency, we also have made all patches generated by NPR techniques and our assessment results available in our replication package. 

\subsubsection{External Validity.} Our research shares a common threat to external validity with prior works~\cite{jiang2021cure, silva2023repairllama, xia2022less, xia2023practical, wei2023copiloting, le2024evaluating} that our results may not generalize across different programming languages and datasets. To mitigate this risk, we evaluated our model, \ours, and state-of-the-art baselines using three datasets, Defects4J~\cite{just2014defects4j}, HumanEval-Java~\cite{jiang2023impact} and TransformedD4J~\cite{le2024evaluating}, and found our performance generalize across these datasets. Additionally, the core contributions of our approach are language-agnostic, suggesting potential applicability to various programming languages. Therefore, we believe that this risk is minimal. 

Another external validity relates to our generalization to larger LLMs such as ChatGPT or Claude models. FLAMES modify the LLM decoding strategy and thus require access to LLMs. This avoids evaluation on such closed-source LLMs. As an alternative, we have done our best to conduct experiments on CodeQwen2.5, the best open source code LLM across leaderboards, and demonstrate the effectiveness of FLAMES on this model in Table~\ref{tab:generalizability}. Moreover, our experiments for RQ5 demonstrate the generalizability of FLAMES across six well-known LLMs with diverse model architectures and sizes. Therefore, we believe that FLAMES is capable of generalizing effectively to different LLMs.

\subsubsection{Construct Validity.} Our methodology may encounter threats to construct validity concerning the appropriateness of our evaluation metrics. Due to limited human resources, besides the number of correct patches, we leverage the number of plausible patches as a proxy effectiveness metric, which might not fully capture the actual effectiveness of APR techniques. To minimize this risk, we only utilized this metric when comparing different variants of the same APR techniques, where we observed consistent trends between correct patches and plausible patches. 

In addition, we also share a common threat to construct validity with previous work~\cite{jiang2023impact, jiang2021cure, chen2019sequencer, wei2023copiloting, xia2022less} on the assumptions of perfect fault localization. However, we did our best efforts to ensure a fair comparison between FLAMES and LLM-based Program Repair by adopting the same fault localization assumptions as our base models as we reused their preprocessing pipelines. For example, in the default setting, our base model, RepairLlama, requires perfect fault localization, so that FLAMES also needs to use perfect fault localization. However, for multi-location bugs, RepairLlama only needs a coarse-grained bug region. As shown in Section~\ref{sec:multi_locations}, FLAMES can effectively repair multi-location bugs under this assumption. These results suggest that FLAMES may operate without perfect fault localization, provided that its base models support such flexibility. 

Finally, as FLAMES relies on test case, it is possible that its performance may not generalize to lower-quality test suite. To address this concern, we evaluated the impact of test suite quality on FLAMES by grouping bugs according to line coverage of their test suites: <60\%, 60–80\%, and >80\%. FLAMES produced plausible patches for 65\%, 69\%, and 62\% of bugs in these groups, and correct patches for 40\%, 42\%, and 40\%, respectively. However, the precision for the >80\% coverage group was slightly higher at 0.64, compared to 0.61 and 0.62 in the lower coverage groups. These results suggest that FLAMES performs consistently across varying test suite quality, with a slight decrease in precision on lower coverage suites.

\section{Related Works}~\label{sec:Related_Work}

\subsection{Large Language Models for Automated Program Repair}

Recent advancements in Large Language Models (LLMs), pre-trained on extensive datasets, have significantly advanced various coding tasks~\cite{fan2023large}, such as code generation~\cite{fried2022incoder, chen2021evaluating}, understanding~\cite{nam2024using, ahmed2022few}, and analysis~\cite{kazerounian2021simtyper, le2022autopruner}. In the field of APR, LLMs have also shown great promise when employed through either fine-tuning techniques~\cite{silva2023repairllama, jiang2023impact, huang2023empirical} or prompting methods~\cite{xia2023practical, xia2023keep}.

Fine-tuning approaches concentrate on refining the weights of code-specific LLMs like CodeLLama~\cite{roziere2023code} and InCoder~\cite{fried2022incoder}, using APR datasets including MegaDiff~\cite{monperrus2021megadiff}. Notably, Jiang et al.\cite{jiang2023impact} demonstrated significant enhancements in LLMs for APR through full-parameter fine-tuning, showing remarkable improvements over the original models. Silver et al.\cite{silva2023repairllama} explored the effectiveness of the QLORA~\cite{dettmers2024qlora} fine-tuning method, assessing various input and output formats in LLM-based APR and introducing RepairLLama, an advanced APR technique using CodeLLama. Jin et al.\cite{jin2023inferfix} introduced InferFix for fixing security vulnerabilities by fine-tuning Codex model and retrieval augmented generation.

Prompting approaches aim to directly leverage LLMs without finetuning. Particularly, they mainly focus on designing effective prompting which provide feedbacks to guide LLMs to generate correct programs. Notably, Xia et al.~\cite{xia2023keep} propose conversation-driven APR, namely ChatRepair, which guide ChatGPT with instant feedbacks  such as test name, relevant test code, and error message, to perform APR in a conversational style. 
Ahmed et al.~\cite{ahmed2023majority} and Yin et al.~\cite{yin2024thinkrepair} levarages chain-of-thought prompting to improve the effectiveness of LLMs on APR. Zhang et al.\cite{zhang2024autocoderover} and Bouzenia et al.\cite{bouzenia2024repairagent} introduce AutoCodeRover and RepairAgent,tools that enable LLMs to follow human-like debugging steps using multi-agent systems to fix real-world issues. 

Different from these studies that typically employ beam search to generate candidate patches, our work empirically investigates the limitations of beam search, particularly when implemented on standard hardware setups. Motivated by these shortcomings, we propose the use of PG-TD to guide LLM-based APR.

\subsection{Efficiency of Automated Program Repair}

In addition to evaluating effectiveness, several studies also focus on the efficiency of APR~\cite{liu2020efficiency, chen2021fast, qi2013efficient, xiao2024accelerating,chen2022program}. Liu et al.~\cite{liu2020efficiency} conducted a systematic assessment of 16 APR techniques, revealing inefficiencies in state-of-the-art APR tools and advocating for further industry exploration of efficiency impacts. Chen et al.~\cite{chen2021fast} introduced UniAPR, a patch validation framework that enhances efficiency by avoiding unnecessary restarts of the JVM for all existing bytecode and source code-level APR techniques. Xiao et al.~\cite{xiao2023expressapr} developed ExpressAPR, which accelerates APR by integrating five mechanisms: mutant schemata, mutant deduplication, test virtualization, test prioritization, and parallelization. Chen et al.~\cite{chen2022program} proposed utilizing XGBoost to develop on-the-fly prioritization techniques to expedite APR.
Kim et al.~\cite{kim2023automated} and Gresino~\cite{kim2024enhancing} presented novel patch-scheduling algorithms for enhancing APR time efficiency.

Different from these studies that typically focus on time efficiency, our approach focus on memory efficiency, motivated by the extensive memory requirements of LLMs. Our work is closely related RepairLLama~\cite{silva2023repairllama}, which explores efficient fine-tuning of LLMs during the training phase. Different from this work, we investigate memory efficiency during the patch generation phase. 
\section{Conclusion}
In this study, we empirically evaluated the impact of beam size on the memory efficiency and effectiveness of LLM-based APR techniques. Our findings reveal that an increase in beam size results in significantly higher VRAM usage. 
Consequently, increasing the beam size causes numerous out-of-memory crashes and subsequently unduly degrades the performance of LLM-based APR. To address this challenge, we introduced \ours—the first APR approach that integrates LLMs with a semantic-guided best-first algorithm to guide the repair process. Our evaluation on various datasets demonstrates that \ours substantially outperforms the state-of-the-art baselines. Additionally, \ours significantly reduces memory consumption by up to 83\% and accelerates the repair process compared to conventional LLM-based APR techniques.

\textbf{Data Availability.} We published replication package including dataset and results and appendix at~\cite{package}.

\section*{Acknowledgement}
This research was supported by The University of Melbourne’s Research Computing Services and the Petascale Campus Initiative.
Thanh Le-Cong is partially supported by Google through its Ph.D. Fellowship program.
Xuan-Bach D. Le is supported by the Australian Government
through the Australian Research Council’s Discovery Early Career
Researcher Award (DECRA) funding

%%
%% The next two lines define the bibliography style to be used, and
%% the bibliography file.
\bibliographystyle{ACM-Reference-Format}
\bibliography{main}

% \newpage
% \appendix

% \input{appendix}
%%
%% If your work has an appendix, this is the place to put it.

\end{document}